\documentclass[preprint]{aastex631}
\usepackage{amsmath}
\usepackage{amssymb}
\newcommand{\Msun}{M_{\odot}}
\newcommand{\unit}[1]{\ensuremath{\, \mathrm{#1}}}
\newcommand{\vbar}{\bar{v}}
\newcommand{\vesc}{v_{esc}}
\newcommand{\Rstar}{R_{\star}}
\newcommand{\Mstar}{M_{\star}}
\newcommand{\be}{\begin{equation}}
\newcommand{\ee}{\end{equation}}
\newcommand{\pn}{p_N(\tau)}
\newcommand{\Ncut}{N_{cut}}
\newcommand{\vn}{v_{N}}

\newcommand{\mess}{\frac{3(\vn^2-\vesc^2)}{2\vbar^2}}
\newcommand{\avg}[1]{\langle #1 \rangle}
\newcommand{\sqsixoverpi}{\left(\frac{6}{\pi}\right)^{1/2}}

\newcommand{\GeV}{\unit{GeV}}

\newcommand{\tenovervbar}{\frac{10~\unit{km}\unit{s}^{-1}}{\vbar}}

\newcommand{\Rsun}{R_{\odot}}

\newcommand{\percc}{\unit{cm}^{-3}}
\shortauthors{Ilie \& Paulin}
\graphicspath{{./}{figures/}}

\begin{document}

\title{Analytic Approximations for the Velocity Suppression of Dark Matter Capture}

\author {Cosmin Ilie}
\affiliation{Colgate University \\
13 Oak Drive \\
Hamilton, NY 13346, USA}
\affiliation{Department of Theoretical Physics, National Institute for Physics and Nuclear Engineering\\
 Magurele, P.O.Box M.G. 6, Romania}
 
\author{Jillian Paulin}
\affiliation{Colgate University \\
13 Oak Drive \\
Hamilton, NY 13346, USA}



\begin{abstract}

Compact astrophysical objects have been considered in the literature as dark matter (DM) probes, via the observational effects of annihilating captured DM. In this paper we investigate the role of stellar velocity on the multiscatter capture rates and find that the capture rates of DM by a star moving with respect to the DM halo rest frame are suppressed by a predictable amount. We develop and validate an analytical expression for the capture rate suppression factor. This suppression factor can be used to directly re-evaluate projected bounds on the DM-nucleon cross section, for any given stellar velocity, as we explicitly show using Population~III stars as DM probes. Those objects (Pop~III stars) are particularly interesting candidates, since they form at high redshifts, in very high DM density environments. We find that previous results, obtained under the assumption of star at rest with respect to the DM rest frame are essentially unchanged, when considering the possible orbital velocities for those central stars.

\end{abstract}



\section{Introduction} \label{sec:intro}

Dark Matter (DM)--- non-baryonic, non-luminous matter that interacts predominantly gravitationally--- has been a scientific puzzle since Zwicky coined the term Dunkele Materie (in translation Dark Matter) in 1933~\citep{Zwicky:1933}.  Today, there are numerous viable theories on the nature of DM. Some of the most notable include: weakly interacting massive particles (WIMPs)~\citep[see][and references therein]{Roszkowski:2017WIMPs}, WIMPZILLAs~\citep{Kolb:1999}, and axions or axion-like particles~\citep[see][and references therein]{Marsh:2015axiions}, to name a few. For a while, massive astrophysical compact halo objects (MACHOs) were also popular candidates~\citep[for a review see][]{Evans:2004MACHOs}, and today a related version of this line of reasoning are Primordial Black Holes (PBHs) as DM candidates~\citep[see][and references therein]{Carr:2020PBHs}. There are two distinct strategies for DM detection. One is direct detection, based on the interactions between DM and baryonic matter and the minute energy transferred to nuclei by collisions with the omnipresent sea of DM particles within our galaxy, through which the Earth and the Sun travel~\citep[for a recent review see][]{Schumann:2019direct}. Any such DM signal should have a clear annual modulation, as predicted by~\cite{Drukier:1986}. 
Intriguingly, such a signal has been detected by the DAMA experiment starting in 1998~\citep{BERNABEI:1998}, and has persisted for more than two decades with an ever increasing statistical significance~\citep{Bernabei:2018}. It is striking that none of the other direct detection experiments have identified a similar signal. Recently, two experiments (ANAIS and COSINE)  have been set up with the same detector technology (NaI) as the DAMA experiment, and, while preliminary, there is no indication of a statistically significant annual modulation in their data~\citep{Adhikari:2019,Amare:2021}.
Other very sensitive DM direct detection searches include the XENON1T~\citep{Aprile_2012,Aprile:2018,Aprile:2019,Aprile:2019ldm,Aprile:2020} experiment in Gran Sasso, Italy, the PICO experiment located at SNOLAB in Canada~\citep{Amole:2019fdf}, and PandaX-II in China~\citep{Tan_2016}, among others. Despite the fact that these experiments have been running for some time, none of them have yet detected DM directly.

Rather than relying solely on direct detection, one can extract DM parameters for any model based on annihilation signals that could originate from DM dense regions. This, in a nutshell, is the essence of indirect detection techniques. For a review see ~\cite{Feng:2000indirect}. For instance, ~\cite{Freese:2008cap,Iocco:2008cap,Ilie:2019,Ilie:2019Erratum,Ilie:2020BNFa,Ilie:2020PopIIIa} discuss the impact of DM on a Pop~III star's luminosity. Most stars shine below the Eddington luminosity--- that is, the brightest theoretically possible luminosity that preserves hydrostatic equilibrium. However, a star that has accreted enough DM may shine at the Eddington limit, and, as such, a limit on its mass can be placed if we know the DM-proton interaction cross section. Conversely, if Pop~III stars (zero metallicity nuclear powered first stars) are observed, their mere existence implies an upper bound on the cross section. 

 The role of astrophysical objects as potential DM laboratories has been recognized in the literature for a while. For instance the pioneering work on DM capture~\citep{1985ApJ...299..994F, Press:1985,Spergel:1985,Gould:1987resonant, Gould:1988} also deals with potentially observable effects this phenomenon has on our Sun or the Earth. Via collisions with nuclei, or electrons, inside the dense environment of a compact object, such as a star, DM particles traversing it can be slowed down. Some of those will lose enough energy to become gravitationally trapped, and therefore captured, by the object. Subsequently they will sink in toward the center of the star, where DM annihilations can produce energy (or secondary particles) that can have observable effects. 
 
 For dense capturing stars, and/or for heavy DM, typically the DM particle will experience more than one collision per crossing. Using the multiscatter capture formalism~\citep{Gould:1992ApJ,Bramante:2017,Dasgupta:2019juq,Ilie:2020Comment,Bell:2020}, the potential of several classes of objects to constrain properties of DM has been explored in the literature. Below we include a non-exhaustive list of the more recent papers where such effects have been analyzed for: Pop~III stars~\citep{Freese:2008cap,Iocco:2008,Taoso:2008PhRvD,Ilie:2019,Ilie:2020PopIIIa,Ilie:2020BNFa}, Neutron Stars~\citep{Gould:1990, Bertone:2008,Kouvaris:2008,Baryakhtar:2017,Bramante:2017,Raj:2017wrv,Croon:2017zcu,Bell:2018pkk,Chen:2018ohx,Gresham:2018rqo,Acevedo:2019,Bell:2019pyc,Hamaguchi:2019oev,Leroy:2019,Leung:2019,Joglekar:2019,Bell:2020,Bell:2020b,Bell:2020NSSINS,Garani:2020,Genolini:2020,Joglekar:2020,Keung:2020,Kumar:2020,Perez-Garcia:2020}, White Dwarfs~\citep{Moskalenko:2007ApJ,Bertone:2008,Bertolami:2014wua,Bramante:2017,Dasgupta:2019juq,Horowitz:2020axx,Panotopoulos:2020kuo}, and exoplanets~\citep{Leane:2020wob}, and the Earth~\citep{Wilczek:1986, FREESE1986295, Gould:1987resonant, Gould:1992cosmological, Mack:2007}.

One common assumption made in most of the aforementioned studies is that the capturing object is at rest with respect to the DM halo. However, as shown by~\cite{Gould:1987resonant}, for the case of single scattering, the capture rates are suppressed when the effects of stellar velocities are included. The aim of this paper is to generalize the result of~\cite{Gould:1987resonant} and provide an analytic estimation of the suppression coefficient, for the more general case of multiscatter capture of dark matter. 

We end the introduction with a description of the structure of this paper. In Section~\ref{sec:DMCap} we consider the case of zero stellar velocity, and briefly review the DM multiscatter capture formalism~\citep{Bramante:2017,Ilie:2020Comment} and the closed form analytic approximation formulae of~\cite{Ilie:2020Comment,Ilie:2020PopIIIa}. In Section~\ref{sec:AnalyticalCap} we present and validate our main result, the velocity dependent suppression coefficients (see Equations~(\ref{eq:Ieta})-(\ref{eq:I0}) and Figure~\ref{fig:SF_sun_full_analytic}). 
Those could prove to be useful for future research, as using  them, in conjunction with the analytic estimates for the zero velocity capture rates, can bypass the need for a full numeric, computationally expensive, calculation. Moreover, the suppression coefficients (Equations~(\ref{eq:Ieta})-(\ref{eq:I0}))  would allow one to quickly rescale, by simply dividing by the corresponding suppression factor, any bounds on DM-nucleon cross section ($\sigma$) obtained under the assumption of a capturing object at rest, once the velocity of the capturing star is known. Finally, in Section~\ref{sec:Bounds} we revisit the bounds obtained in~\cite{Ilie:2020BNFa,Ilie:2020PopIIIa} by using Pop~III stars as DM probes. Namely, using the formalism developed in Section~\ref{sec:AnalyticalCap} we estimate the role of a possible stellar velocity on the projected bounds on $\sigma$ and find that for Pop~III stars stellar velocity only weakens the bounds by at most a factor of a few. We end with Section~\ref{sec:Conclusions}, where our conclusions are presented. 

We want to emphasize that our main results, presented in Section~\ref{sec:AnalyticalCap} can be applied to any astrophysical object that is capturing Dark Matter, and are not restricted to the Sun or Pop~III stars, which were the focus of this paper. The main reason for us restricting our attention to Pop~III stars in Section~\ref{sec:Bounds}, where we estimate projected bounds on $\sigma$, is that our main motivation for this work was to re-evaluate, as explained above, the forecast bounds previously obtained by our group.

\section{DM Capture by objects at rest} \label{sec:DMCap}

In this section we give a brief overview of the formalism necessary to predict the capture rates of DM by astrophysical compact objects, such as stars, planets, etc. In order for a DM particle to be captured by a star, its velocity must fall below the star's escape velocity. This can occur through collisions with baryonic nuclei in the star. Depending on the mass of the DM particle, this may happen after one~\citep{Press:1985,Gould:1988,Gould:1987resonant} or more collisions~\citep{Gould:1992ApJ,Bramante:2017,Dasgupta:2019juq,Bell:2020,Ilie:2020Comment}, where very massive DM particles will need more collisions for capture than less massive DM particles. Additionally, the number of collisions that are likely to occur depends on the characteristics of the star and the cross section of interaction between DM and targets inside the star; this number is roughly equal to the optical depth, $\tau=n_T\sigma(2R_{\star})$, where $n_T$ is the average number density of target particles in the star, $\sigma$ is the cross section of interaction, and $R_{\star}$ is the stellar radius. For all objects considered we will assume one constituent dominates over all others. For example,in the case of Pop~III stars, or the Sun, which are composed primarily of hydrogen, we assume the atomic nuclei to have the mass of one proton. The probability of capture after exactly $N$ scatters may be represented as follows~\citep{Bramante:2017}:

\begin{equation}\label{eq:CN}
    C_N= \pi R_{\star}^2 p_N(\tau) \int_{0}^{\infty} f(u) \frac{du}{u} w^2 g_N (w),
\end{equation}
 where $R_{\star}$ is the radius of the star, $p_N(\tau)$ is the probability of $N$ collisions occurring, $f(u)$ is the DM velocity distribution, and $g_N(w)$ is the probability that the DM particle's velocity will fall below the escape velocity after $N$ scatters. The quantity $p_N(\tau)$ may be represented as~\citep{Ilie:2019}:

\begin{equation}\label{eq:pN}
    \pn=\frac{2}{\tau^2}\left(N+1-\frac{\Gamma(N+2,\tau)}{N!}\right),
\end{equation}
with $\Gamma(a,b)$ is the upper incomplete gamma function defined as $\Gamma(a,b)=\int_b^\infty t^{a-1}e^{-t}\,dt$. As found in~\cite{Bramante:2017}, $g_N(w)$ can be approximated with:

\begin{equation}\label{eq:gN}
    g_N(w)=\Theta (v_{esc}(1-\frac{\beta_{+}}{2})^{-N/2}-w),
\end{equation}
where $\beta_{+}=\frac{4m_{\chi} m_T}{\left(m_{\chi} + m_T \right)^2}$, with $m_{\chi}$ being the DM particle mass, $m_T$ the mass of the target particle, and $\vesc$ is the escape velocity at the surface of the star. Additionally, $w$ represents the velocity of a DM particle as it enters the star, and is related to its velocity infinitely far away ($u$) by $w^2=v_{esc}^2+u^2$. This last statement is just conservation of energy. In order to determine the total capture rate, we must sum the values of $C_N$ for every value of $N$:

\begin{equation}\label{eq:Ctot}
    C_{tot}=\sum_{N=1}^{\infty} C_N .
\end{equation}
This is a complete analytical representation of the total capture rate. However, in order to perform a numerical calculation, it is impossible to sum to infinity. Therefore it is necessary to implement a cutoff condition. We continue summing the series up to $\Ncut$, when we reach a desired level of accuracy which we arbitrarily set to $0.1\%$; that is, until one additional iteration of $C_N$ only changes $C_{tot}$ by $0.1\%$. As shown in~\cite{Ilie:2019}, convergence is attained when $\Ncut\sim\tau$, i.e., whenever we sum up to the average number of collisions a DM particle experiences, per crossing, with targets inside the star.

Next we restrict our attention to a capturing object at rest with respect to the DM halo rest frame. In this situation, the DM velocity distribution $f(u)$ is simply the Maxwell-Boltzmann distribution $f_0$~\citep{Gould:1988}:
\begin{equation}\label{eq:MBdist}
    f_0(u)du=n_{\chi}\frac{4}{\sqrt{\pi}}x^2\exp(-x^2)dx,
\end{equation}
where $n_{\chi}$ is the number density of DM particles, $x$ is a dimensionless quantity defined as $x\equiv\sqrt{\frac{m_\chi}{2T_{\chi}}}u$, with $T_{\chi}$ representing the DM temperature, which can be related to the thermal average velocity of DM particles: $\bar{v}\equiv\sqrt{3T_{\chi}/m_{\chi}}$. For this case the integral representation $C_N$ presented in Equation~(\ref{eq:CN}) has a closed form analytic solution~\citep{Bramante:2017,Ilie:2020Comment}:

\be\label{eq:CNClosed}
C_{N}=\frac{1}{3}\pi \Rstar^{2} p_{N}(\tau) \frac{\sqrt{6} n_{\chi}}{\sqrt{\pi} \bar{v}}\left(\left(2 \bar{v}^{2}+3 v_{e s c}^{2}\right)-\left(2 \bar{v}^{2}+3 v_{N}^{2}\right) \exp \left(-\frac{3\left(v_{N}^{2}-v_{e s c}^{2}\right)}{2 \bar{v}^{2}}\right)\right),
\ee
where $v_N=\vesc(1-\avg{z}\beta_+)^{-N/2}$, with $\avg{z}$, the average of the kinematic variable defined that accounts for the scatter angle, and for which a good approximation is $\avg{z}\approx 1/2$~\citep{Bramante:2017}. 

Several useful analytic approximations for the total capture rates based on summing the $C_N$s of Equation~(\ref{eq:CNClosed}) have been derived in ~\cite{Ilie:2020Comment,Ilie:2020PopIIIa}. We reproduce those results here, for convenience, and future reference. 

\begin{equation}
C_{tot}\approx\begin{cases}
\left(\frac{2}{3\pi}\right)^{1/2}\frac{\pi \Rstar^{2}}{\tau^{2}}n_{\chi}\frac{3\vesc^{2}+2\vbar^{2}}{\vbar}\Ncut(\Ncut+3), & \text{if }~ R_{v}\ll 1\\
\sqsixoverpi\frac{\pi \Rstar^{2}}{\tau^{2}}n_{\chi}\frac{\vesc^{4}}{\vbar^3}\beta_{+}\avg{z}\Ncut(\Ncut+1)(\Ncut+2)\left(1+\frac{\beta_{+}\avg{z}}{4}(1+3\Ncut)\right),\! &~\text{if }~R_{v}\gg 1,
\end{cases}
\end{equation}

where we defined $R_v\equiv\mess$, and, as pointed out before, the series defining $C_{tot}$ converges at $\Ncut\sim\tau$.

For most astrophysical objects of interest (with Earth being an important exception), the escape velocity is much larger than the thermal velocity of dark matter($\vesc \gg \vbar $). Assuming there is a definite hierarchy between $m_\chi$ and $m_T$, i.e., if $m_\chi\gg m_T$ or $m_\chi\ll m_T$ Equation~(\ref{eq:CNClosed}) could be simplified as: 
\be\label{eq:CNapprox}
C_{N} = \sqrt{24\pi}n_{\chi}GM_{\star}R_{\star}\frac{1}{\bar{v}} p_{N}(\tau)\left(1-\left(1+\frac{2 A_{N}^{2} \bar{v}^{2}}{3v_{esc}^{2}}\right) e^{-A_{N}^{2}}\right), \text{where} \  A_{N}^{2}=Nk,
\ee
where we defined the following dimensionless parameter:
\be\label{eq:Defk}
k\equiv3\frac{\min(m_T;m_\chi)}{\max(m_T;m_\chi)}\frac{\vesc^2}{\vbar^2}. 
\ee
Using the approximate form of $C_N$ from Equation~(\ref{eq:CNapprox}) in~\cite{Ilie:2020Comment,Ilie:2020PopIIIa} we found closed form approximations of the total capture rate, which we reproduce below~\footnote{For more details, derivations, and numerical validations see~\cite{Ilie:2020PopIIIa}.}. Those approximations are functionally different, depending on the region of the $\sigma-m_{\chi}$ parameter space. For the case of multiscattering capture ($\tau\gtrsim1$) we find two distinct regimes. First, in the region we called Region~I ($\tau\gtrsim 1$ and $k\tau\lesssim 1$):
\be\label{eq:CtotScalI}
C_{tot}^{I}\approx5\times10^{54}~\unit{s}^{-1}\left(\frac{\rho_{\chi}\times\sigma}{\GeV~\unit{cm}^{-1}}\right)\left(\frac{10^8~\GeV}{m_{\chi}}\right)^2\left(\tenovervbar\right)^3\left(\frac{\Mstar}{\Msun}\right)^3\left(\frac{\Rstar}{\Rsun}\right)^{-2}.
\ee
In what we called Region~II, defined $\tau\gtrsim 1$ (multiscatter) and $k\tau\gtrsim 1$, we find that the capture rates are insensitive to the cross section $\sigma$. This essentially means that the cross section is so high~\footnote{In the literature this $m_{\chi}$ dependent cross section is called the ``geometric cross section.''} once we cross the boundary between regions~I and~II, that the entire DM flux crossing the object gets captured whenever $\sigma$ is in Region~II of the parameter space, and the capture rate saturates:
\be\label{eq:CtotScalII}
C_{tot}^{II}\approx 8\times 10^{29}~\unit{s}^{-1}\left(\frac{\rho_\chi}{\GeV~\unit{cm}^{-3}}\right)\left(\frac{10^2~\GeV}{m_\chi}\right)\left(\tenovervbar\right)\frac{\Mstar}{\Msun}\frac{\Rstar}{\Rsun}.
\ee
Moving on to the single scattering regime ($\tau\lesssim 1$) we find two distinct functional forms of the capture rates, depending on the relative size of the parameter $k$ when compared to unity. In what we called Region~III, i.e., $\tau\lesssim 1$ and $k\gtrsim 1$, we find:
\be\label{eq:CtotScalIII}
C_{tot}^{III}\approx 4.3\times 10^{64}~\unit{s}^{-1}\left(\frac{\rho_{\chi}\times\sigma}{\GeV~\unit{cm}^{-1}}\right)\left(\frac{10^2~\GeV}{m_\chi}\right)\left(\tenovervbar\right)\left(\frac{\Mstar}{\Msun}\right)^2\left(\frac{\Rstar}{\Rsun}\right)^{-1}.
\ee
Finally, in Region~IV, defined as $\tau\lesssim 1$ and $k\lesssim 1$ we find, remarkably, that the capture rate has the exact same parametric form as that of Region~I ($\tau\gtrsim 1$ and $k\tau\lesssim 1$):
\be\label{eq:CtotScalIV}
C_{tot}^{IV}\approx5\times10^{54}~\unit{s}^{-1}\left(\frac{\rho_{\chi}\times\sigma}{\GeV~\unit{cm}^{-1}}\right)\left(\frac{10^8~\GeV}{m_{\chi}}\right)^2\left(\tenovervbar\right)^3\left(\frac{\Mstar}{\Msun}\right)^3\left(\frac{\Rstar}{\Rsun}\right)^{-2}.
\ee

In summary, in this section we have briefly reviewed the multiscatter DM capture formalism of~\cite{Bramante:2017}. Applying it to the case of zero stellar velocities, we reproduced useful closed form analytic formulae for the total capture rates, previously obtained in the literature~\citep{Ilie:2020Comment,Ilie:2020PopIIIa}. 

In the next section we move to the main aim of our paper, that of addressing the following question: is it possible obtain similar analytic, closed form formulae for the total capture rates when considering the more general case of a capturing object moving with respect to the DM halo rest frame? As we will show shortly, the answer is yes. This could prove to be useful for future research, as full numeric calculations are computationally expensive, especially when coupling DM capture to stellar evolution codes in order to assess the effects of captured DM annihilations on the stellar structure and evolution.

\section{Analytical Evaluation of the velocity suppressed DM Capture Rate} \label{sec:AnalyticalCap}

In this section we present an analytical approximation of the suppression factor for the DM capture rates, in both the low (i.e., $k\gg1$) and high (i.e., $k\ll1$) DM mass regimes. This can be very useful when one needs to estimate the effects of the stellar velocity on DM capture rates, and implicitly on DM scattering cross section bounds, since calculating numerically the capture rates including the full, boosted MB distribution can be quite computationally expensive. Our procedure allows one to calculate the simpler, and fully analytically solvable~\citep{Ilie:2020PopIIIa} rates when the stellar velocity is neglected, and then apply the suppression factor we derive for any given $\eta$. Such a procedure is quite useful when considering capture of DM by astrophysical probes within the Solar System neighborhood, where, based on DM profile \citep{Lin:2019b} and dispersion velocity \citep{Brown:2009} estimates, we would expect $\eta$ to be on the order of a few. 

In principle, the capture rates of DM by astrophysical objects that have a non-zero velocity with respect to the DM halo rest frame are straightforward to calculate numerically. Essentially $C_{tot}$ is still a series obtained by summing the partial capture rates $C_N$, as given by Equation~(\ref{eq:Ctot}). The only change now is that, when calculating  each $C_N$ numerically via the integral over the DM distribution given in Equation~(\ref{eq:CN}), we need to use the appropriate DM distribution. As shown in~\cite{Gould:1987resonant}, this ``boosted'' distribution ($f_{\eta}$) can be easily related to the Maxwell-Boltzmann distribution ($f_0$, see Equation~(\ref{eq:MBdist})) that would be appropriate to use when the star is stationary:

\begin{equation}\label{eq:boostMB}
    f_\eta (u) = f_0 (u) \exp(-\eta^2) \frac{\sinh\left(2x\eta\right)}{2x\eta},
\end{equation}
with $f_0(u)$ given in Equation~(\ref{eq:MBdist}). The parameter $\eta$ represents the dimensionless stellar velocity $\tilde{v}$, normalized to the dispersion velocity of DM particles in the halo:
\begin{equation}\label{eq:eta}
    \eta \equiv \sqrt{\frac{3}{2}}\frac{\tilde{v}}{\bar{v}}.
\end{equation}
Rather than integrating numerically over the DM velocity distribution to calculate a ``boosted'' capture rate, we can find an equivalent analytical expression. In order to develop this, we followed a similar method to~\cite{Bramante:2017,Ilie:2020Comment}. The main difference comes from the use of the boosted velocity distribution as outlined by~\cite{Gould:1987resonant} instead of the assumption of a Maxwell-Boltzmann distribution. Evaluating Equation~(\ref{eq:CN}) by substituting Equation~(\ref{eq:boostMB}) for $f(u)$, we obtain:
\be\label{eq:CNboost}
\begin{split}
    C_N = & \frac{n_{\chi}\pi p_N(\tau)R^2}{2\sqrt{6}\bar{v} \eta}\Bigg[ \frac{1}{\sqrt{\pi}} \exp{\left(\frac{-3(v_N^2-v_{esc}^2)}{\bar{v}^2}-2\eta^2\right)}\bar{v}\Bigg\{4\exp{\left(\frac{3(v_N^2-v_{esc}^2)}{\bar{v}^2}+\eta^2\right)}\bar{v}\eta+\\
    &\exp{\left(\frac{-6v_{esc}^2+6v_N^2-4\sqrt{6}v_{esc}\bar{v}\eta\sqrt{-1+\frac{v_N^2}{v_{esc}^2}}+4\bar{v}^2\eta^2}{4\bar{v}^2}\right)}\left(\sqrt{6}v_{esc}\sqrt{-1+\frac{v_N^2}{v_{esc}^2}}-2\bar{v}\eta\right)- \\
    &\exp{\left(\frac{\sqrt{\frac{3}{2}}v_{esc}\sqrt{-1+\frac{v_N^2}{v_{esc}^2}}}{\bar{v}}+\eta\right)^2}\left(\sqrt{6}v_{esc}\sqrt{-1+\frac{v_N^2}{v_{esc}^2}}+2\bar{v}\eta\right)\Bigg\}+ \\
    &\left(3v_{esc}^2+\bar{v}^2\left(1+2\eta^2\right)\right)\text{erf}{\left(\frac{\sqrt{\frac{3}{2}}v_{esc}\sqrt{-1+\frac{v_N^2}{v_{esc}^2}}}{\bar{v}}-\eta\right)}+
    2\left(3v_{esc}^2+\bar{v}^2\left(1+2\eta^2\right)\right)\text{erf}\left(\eta\right)- \\
    &\left(3v_{esc}^2+\bar{v}^2\left(1+2\eta^2\right)\right)\text{erf}\left(\frac{\sqrt{\frac{3}{2}}v_{esc}\sqrt{-1+\frac{v_N^2}{v_{esc}^2}}}{\bar{v}}+\eta\right)\Bigg],
\end{split}
\ee

 where the quantity $v_N$ is defined as follows~\citep{Bramante:2017}:
 \begin{equation}\label{eq:vN}
    v_N=v_{esc}\left(1-\frac{\beta_{+}}{2}\right)^{-\frac{N}{2}}= v_{esc}\bigg|{\frac{m_\chi^2 + m_T^2} {(m_\chi + m_T)^2}}\bigg|^{-N/2}.
 \end{equation}
 As a sanity check, we verify that in the limit of $\eta=0$, the expression reduces to that found in Equation~(\ref{eq:CNClosed}):
 \begin{equation}\label{eq:CNboost_limit_eta0}
    \lim_{\eta \to 0} C_N = \sqrt{\frac{2\pi}{3}} \frac{n_{\chi} p_N(\tau) R^2}{\bar{v}} \left[(2\bar{v}^2+3v_{esc}^2)-\exp\left({\frac{-3(v_N^2-v_{esc}^2)}{2\bar{v}^2}}\right)(3v_{N}^2+2\bar{v}^2)\right],
 \end{equation}
which corresponds to the case of zero stellar velocity, i.e., $\eta=0$. The expression of $C_N$ presented in Equation~(\ref{eq:CNboost}) is not particularly illuminating. However, it could be used to evaluate the total capture rate as a series, by adding each $C_N$ from $N=1$ until the series has reached the desired level of convergence. This would avoid a full numeric calculation, where each $C_N$ is obtained via Equation~(\ref{eq:CN}). Below we present an alternative to this procedure, based on a predictable ratio between the total capture rates when stellar velocities are non zero and the total capture rate when the stellar velocity is zero, with all other parameters kept the same. A velocity dependent suppression of the capture rates, for the case of single scattering was a generic result found in~\cite{Gould:1987resonant}, and simple analytic estimates were provided for two limiting regimes, high and low DM mass (see Equation~(2.30) of~\cite{Gould:1987resonant}). In this paper we provide a full analytic, exact, closed form solution for the suppression factor, valid in the single scattering regime. Additionally, we generalize this to the case of multiscatter capture of DM. Keeping the same notation as~\cite{Gould:1987resonant}, we define the ``Suppression Factor'', labeled as $\xi_{\eta}$, as the ratio between the total capture rate when $\eta$ is non-zero to the total capture rate for $\eta=0$:
\begin{equation}\label{eq:SupFactDef}
    \xi_{\eta}\equiv\frac{C_{tot}(\eta)}{C_{tot}(\eta=0)}
 \end{equation}
 We next  estimate an upper bound and a lower bound on the suppression factor defined above, and identify the conditions under which those two are equal, and as such equal to $\xi_{\eta}$ itself. We start by noting that the role of the function $g_N(w)$ in the integrals defining $C_N$, and correspondingly in the definition of the suppression factor, is to impose a cutoff on the integral. Namely, this amounts to accounting for the DM particles in the tail of the DM distribution that  are too fast to be slowed down and captured after $N$ collisions. From Equation~(\ref{eq:gN}) one can show that:
\be\label{eq:umax}
u_{max}(N)=v_{esc}((1-\frac{\beta_{+}}{2})^{-N}-1)^{\frac{1}{2}}.
\ee
Defining $a\equiv\min{\{u_{max}(N)\}}$ and $b\equiv\max{\{u_{max}(N)\}}$, for future convenience, and by combining Equations~(\ref{eq:CN}) and~(\ref{eq:Ctot}) and one can show that  $\xi_{\eta}$ lies between a lower and and upper limit given by:
\begin{equation}\label{eq:SupFactULB}
\frac{\int_{0}^{a} f_{\eta}(u) \frac{du}{u} w^2}{\int_{0}^{b} f_{0}(u) \frac{du}{u} w^2}\leq\xi_{\eta}\leq\frac{\int_{0}^{b} f_{\eta}(u) \frac{du}{u} w^2}{\int_{0}^{a} f_{0}(u) \frac{du}{u} w^2}. 
\end{equation}
Whenever $a=b$ then the upper bound and the lower bound on the suppression factor~($\xi_{\eta}$) coincide, and, moreover, they are equal to the suppression factor itself. For the single scattering case, this happens naturally, since $N=1$ so there is only one term in the series $\{u_{max}(N)\}$. Below we provide analytic formulae for the bounds limiting the suppression factor and investigate in detail the conditions under which this can be approximated, not only constrained, in the multiscatter capture case. Changing variables to the dimensionless $x^2\equiv\frac{3 u^2}{2\vbar^2}$, and introducing the following convenient notations:
 \begin{align}
     I_{\eta}(t)&\equiv \exp \left(-\eta ^2\right) \int_0^{t}\exp \left(-x^2\right) \left(\frac{3 x v_{esc}^2}{2 v^2}+x^3\right) \frac{\sinh (2 \eta  x)}{2 \eta  x} \, dx, \label{eq:DefIeta}\\
     I_{0}(t)&\equiv \int_0^{t} \exp \left(-x^2\right) \left(\frac{3 x v_{esc}^2}{2 v^2}+x^3\right) \, dx , \label{eq:IMB}
 \end{align}
one can show that the lower  and upper bounds of the suppression factor can be recast as:
\begin{align}
    \xi_{\eta}^{L.B.}&=\frac{I_\eta(x(a))}{I_{0}(x(b))},\label{eq:SupFactLB}\\
    \xi_{\eta}^{U.B.}&=\frac{I_\eta(x(b))}{I_{0}(x(a))},\label{eq:SupFactUB}
\end{align}
with  $x(a)=\sqrt{\frac{3 a^2}{2\vbar^2}}$, and  $x(b)=\sqrt{\frac{3 b^2}{2\vbar^2}}$, where $a$ and $b$ are the minimum, and respectively maximum of the sequence  $\{u_{max}(N)\}$, defined in Equation~(\ref{eq:umax}). Note that $I_0(t)$ is just $I_\eta(t)$, in the limit of $\eta=0$. For the generic integral $I_\eta(t)$ we find the following closed form solution:
\be\label{eq:Ieta}
I_\eta(t)=\frac{e^{-\eta^2}}{16\eta}\left(4\eta+2 \sqrt{\pi } e^{\eta ^2} \left(2 \eta ^2+1\right) \text{erf}(\eta )+A(\eta;t)+B(\eta;t)\right),
\ee
with:
\be\label{eq:Aeta}
    A(\eta;t)\equiv e^{-t (2 \eta +t)} \left[\sqrt{\pi } \left(2 \eta ^2+1\right) e^{(\eta +t)^2} \left(\text{erf}(t-\eta )-\text{erf}(\eta +t)\right)-2 \eta -2 e^{4 \eta  t} (\eta +t)+2 t\right],
\ee
and
\be\label{eq:Beta}
B(\eta;t)\equiv\frac{3 \sqrt{\pi } e^{\eta ^2} v_{\text{esc}}^2 (2 \text{erf}(\eta )+\text{erf}(t-\eta )-\text{erf}(\eta +t))}{\bar{v}^2}.
\ee
Equations~(\ref{eq:SupFactLB})-(\ref{eq:Beta}) can be used to compute exactly the lower and upper bounds for the suppression factor. Moreover, for single scattering capture (i.e., $\tau\ll 1$) the same set of equations predict exactly the value of the suppression factor, since $x(a)=x(b)=\sqrt{\frac{3u_{max}(1)^2}{2\vbar^2}}$. To gain further insight it is instructive to take limiting cases. First, we explicitly write  the analytic form we find for $I_0(t)$:
\be\label{eq:I0}
I_0(t)=\frac{1}{2} \left(\frac{3 e^{-t^2} \left(e^{t^2}-1\right) v_{\text{esc}}^2}{2\bar{v}^2}- e^{-t^2} \left(t^2+1\right)+1\right)
\ee

We will next restrict our attention to the case when there is a definite hierarchy between $m_\chi$ and $m_T$, i.e., when one of those mass scales is larger than the other. In this case the parameter $\beta_+$ is much less than unity, and can be approximated as: 
\be\label{eq:betaPlapprox}
\beta_+\approx 4\frac{\min(m_T;m_\chi)}{\max(m_T;m_\chi)}.
\ee
We can now approximate the terms in the sequence $\{u_{max}(N)\}$, defined by Equation~(\ref{eq:umax}):
\be\label{eq:umaxapprox}
u_{max}(N)\approx\vesc\left(N\beta_+/2\right)^{1/2}.
\ee
The last two equations combined with the definition of $k$ from Equation~(\ref{eq:Defk}) can be used to show that: $x(a)\approx \sqrt{k}$ and $x(b)\approx \sqrt{k\max(\tau;1)}$.
In the last step we used the fact that for multiscatter capture $N_{max}\approx \tau$. In order to keep the treatment of single scatter and multiscatter unified we used $\max(\tau;1)$, since $N_{max}=N=1$ for single scattering. However, as discussed before, for single scatter the upper and lower bounds of Equations~(\ref{eq:SupFactLB})-(\ref{eq:SupFactUB}) coincide, since in that case $x(a)=x(b)=\sqrt{k}$. We are now in position to derive limiting cases of the lower and upper bounds of the suppression factor of Equations~(\ref{eq:SupFactLB})-(\ref{eq:SupFactUB}). We start with the case of single scattering capture, for which there are two natural regimes: the $k\gg 1$ regime and the $k\ll1$ regime. From the definition of $k$ in Equation~(\ref{eq:Defk}) one can find that $k\gg 1$ is valid whenever $ (3\vesc^2/\vbar^2)^{-1}m_T\ll m_\chi\ll  (3\vesc^2/\vbar^2)m_T$, whereas $k\ll 1$ otherwise. In the $k\gg1$ limit  (corresponding to low $m_\chi$) we perform an asymptotic expansion of $I_\eta$ and $I_0$ from Equations~(\ref{eq:Ieta})-(\ref{eq:I0}) around $x(a)=x(b)=\sqrt{k}\to\infty$. Keeping only leading order terms, we get:

 \be\label{eq:supfactlowmxSS}
     \xi_\eta^{k\gg1}\approx\frac{e^{-\eta^2}\left(4\eta + \frac{6e^{\eta^2}\sqrt{\pi}v_{esc}^2\text{erf}\left(\eta\right)}{\bar{v}^2}+2e^{\eta^2}\sqrt{\pi}\left(1+2\eta^2\right)\text{erf}\left(\eta\right)\right)}{4\left(2+ \frac{3v_{esc}^2}{\bar{v}^2}\right)\eta}.
\ee
Whenever $\vesc^2\gg\vbar^2$, and $\eta$ is not much larger than unity we can further simplify the previous result to:
$\xi_\eta^{k\gg1}\approx\frac{\sqrt{\pi } \text{erf}(\eta )}{2 \eta }$, which matches the result of Gould~\citep[see Equation 2.30 of][]{Gould:1987resonant}.

While still in the single scatter regime, at either very high or very low $m_\chi$, the parameter $k$ becomes much less than unity. Therefore we simply Taylor expand $I_\eta$ and $I_0$ from Equations~(\ref{eq:Ieta})-(\ref{eq:IMB}) around $x(a)=x(b)=\sqrt{k}\approx 0$. Neglecting terms of $\mathcal{O}(k^3)$, we find that the suppression factor $\xi_\eta$ can be approximated with:
\be\label{eq:supfacthighmxSS}
\xi_\eta^{k\ll1}\approx e^{-\eta ^2}+\frac{1}{3} e^{-\eta ^2} \eta ^2 k+\frac{e^{-\eta ^2} \eta ^2 k^2 \left(10 \bar{v}^2+3 \left(4 \eta ^2-5\right) v_{\text{esc}}^2\right)}{270 v_{\text{esc}}^2}.
\ee
We emphasise once more that, for the single scattering regime, the full, non-approximated, functional form of the suppression factor that can be obtained from $\xi_\eta=I_\eta(x(a))/I_0(x(a))$, with $x(a)=\sqrt{3u_{max}^2(1)/2\vbar^2}$, and $I_\eta$ given in Equation~(\ref{eq:Ieta}) and $I_0$ from Equation~(\ref{eq:I0}). However, the approximations derived above allow one to gain some additional insight. In the low $m_\chi$ regime, defined by the $k\gg 1$ condition, the suppression factor has a roughly constant value, given by Equation~(\ref{eq:supfactlowmxSS}). Once $m_\chi$ becomes either extremely low, or extremely large, such that $k$ crosses unity, and now becomes less than one, the suppression factor starts to change significantly, in an approximately polynomial fashion, according to Equation~(\ref{eq:supfacthighmxSS}). Whenever $k$ becomes much less than unity, the suppression factor asymptotes to $e^{-\eta ^2}$, {matching the result found by Gould~\citep[see Equation 2.30 of][]{Gould:1987resonant}. We note here that $k\ll 1$ is equivalent to $\vbar\gg\sqrt{3\frac{\min(m_T;m_\chi)}{\max(m_T;m_\chi)}}\vesc$. Therefore, whenever the DM dispersion velocity is high compared to the escape velocity, only a small fraction of the DM particles will be captured, as most of them will have speeds larger than the escape velocity. 

We next move our focus to the multiscatter capture case. For all objects we considered, it turns out that $k\tau\ll 1$, given the present bounds on $\sigma$ from direct detection experiments. In turn, that means that for the multiscatter case $x(b)\approx \sqrt{k\tau}\ll 1$. Moreover, the same bounds on $\sigma$ imply that $x(a)\approx \sqrt{k}\ll 1$, whenever $\tau\gg1$, i.e., in the multiscatter regime. Expanding around $x(a)\approx \sqrt{k}\approx 0$ and  $x(b)\approx \sqrt{k\tau}\approx 0$ we get the following approximations for the lower and upper bounds of $\xi_\eta$, defined in Equations~(\ref{eq:SupFactLB})-(\ref{eq:SupFactUB}):
\begin{align}
    \xi_\eta^{L.B.}\approx&\frac{e^{-\eta ^2}}{\tau}\\
    \xi_\eta^{U.B.}\approx& \min(e^{-\eta ^2}\tau;1).
\end{align}
While those bounds can be useful, it turns out that in most cases of interest the suppression factor itself can be well approximated with $\xi_\eta\approx e^{-\eta^2}$, as shown below.  Each of the $C_N$ in the definition of the total capture rate ($C_{tot}$) is defined as per Equation~(\ref{eq:CN}). Under the conditions we explore here, i.e., when there is a mass hierarchy between $m_\chi$ and $m_T$, and using the integrals $I_\eta$ defined in Equations~(\ref{eq:Ieta})-(\ref{eq:IMB}) one can show that:
\be
C_N\sim \pn I_\eta(kN),
\ee
up to constants independent of $N$. For the case of $\eta=0$ we have $C_N(\eta=0)\sim\pn I_0(kN)$. As explained above, bounds on $\sigma$ from direct detection experiments imply that, in most cases of interest, $k\ll1$ and $k\Ncut\ll1$, once $\tau\gg1$, i.e., for multiscatter capture. Therefore we can expand both $I_\eta$ and $I_0$ around zero. Keeping leading order terms and we have, up to constants independent of $N$, the following scaling relations: $C_N\sim e^{-\eta^2}Nk$, and $C_N(\eta=0)\sim Nk$. It is important to note that in both terms the same constants were ``ignored.'' As such, the suppression factor in the multiscatter regime becomes simply $\xi_\eta^{M.S.}\approx e^{-\eta^2}$. Note that this is a smooth continuation of the asymptotic behavior found in the single scattering regime (see discussion in the paragraph following Equation~(\ref{eq:supfacthighmxSS})). 

In Figure~\ref{fig:SF_sun_full_analytic}  we validate our analytic results for the suppression factor, against the full numeric result, using the Sun as a sample capturing object. In order to explicitly show the dependence of the suppression factor on the stellar velocity, in Figure~\ref{fig:SF_sunlike_full_analytic} we consider a sun-like star for which we arbitrarily set $\eta=5$, with all other parameters being fixed. Note the significant suppression in this case, when contrasted to an object in the Solar System neighborhood, where $\eta\approx 1$. {We point out that for the Milky Way, as demonstrated by observed rotation curves, the value of the stellar velocities, and in turn the value of $\eta$, is roughly constant, for stars farther than a few kiloparsecs from the galactic center. As such, our choice of $\eta=5$ should be viewed as a hypotetical example, only for the purpose of illustrating how rapidly the exponential suppression factor can reduce capture rates, even for order unity values of $\eta$.}

 \begin{figure}
     \centering
     \includegraphics[scale=0.99]{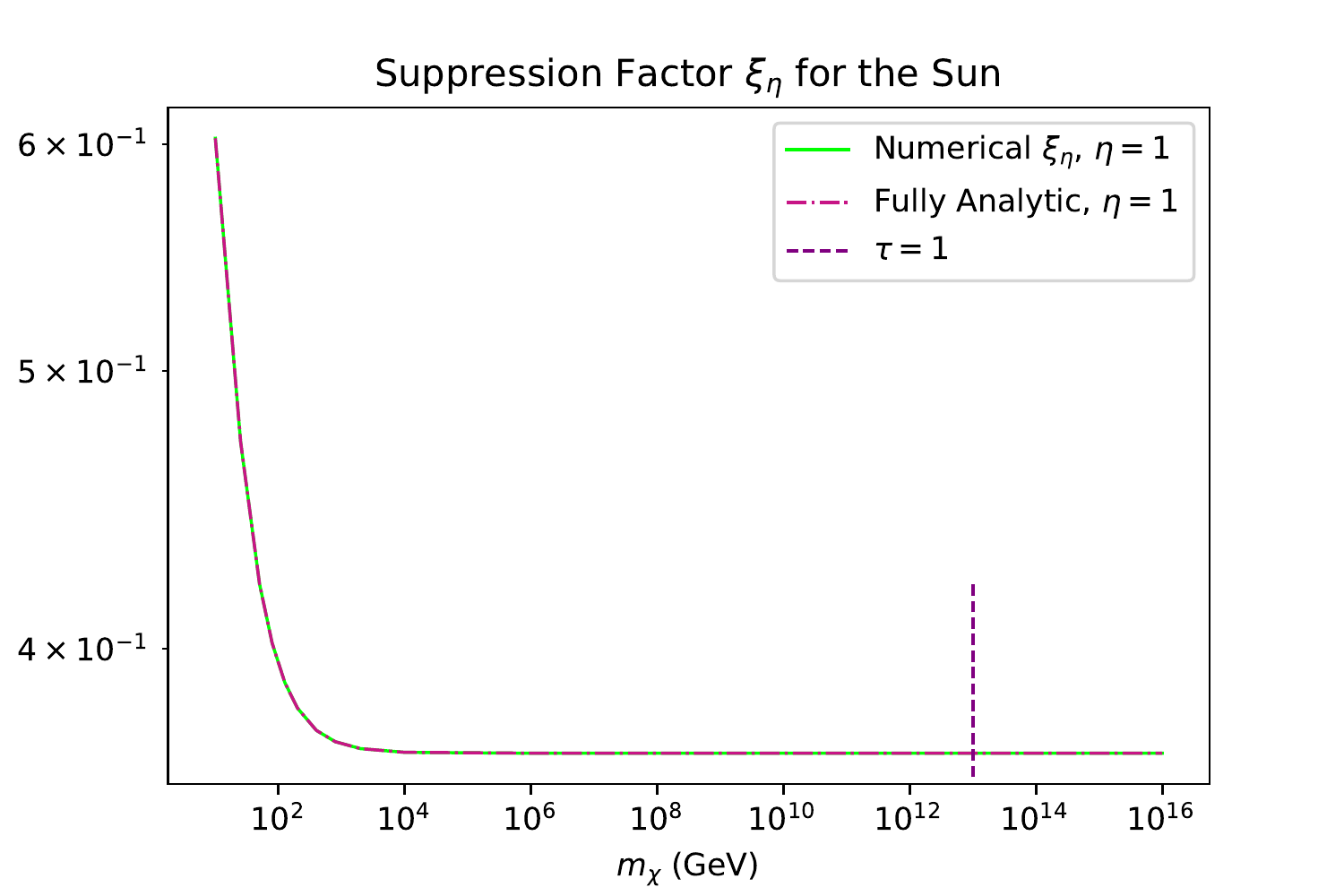}
     \caption{The suppression factor $\xi_{\eta}$ for the Sun calculated numerically (solid green line) and using our fully analytic method via Equations~(\ref{eq:SupFactLB})-(~\ref{eq:I0}) (dash-dotted pink line). Note the full agreement between those two procedures. We used $\eta=1$, $\rho_{\chi}=1~\GeV\percc$, and $\bar{v}=2.2\times10^7$ cm/s. Additionally, the dashed purple line at $m_\chi\sim 10^{13}~\GeV$ marks the  mass where the transition to multiscatter capture happens, for the Sun, if $\sigma$ is assumed at the deepest constraints for $\sigma$ vs $m_\chi$ from Xenon 1T~\citep{Aprile:2020}. We point out that the suppression factor is in fact $\sigma$ independent; however, when calculating it numerically we need to assume a value for $\sigma$. Moreover, the reason we plot the transition from single ($m_\chi\lesssim 10^{13}~\GeV$) to multiscatter capture  ($m_\chi\gtrsim 10^{13}~\GeV$) is to show explicitly that our analytic formalism is valid on both of those regimes.}
     \label{fig:SF_sun_full_analytic}
 \end{figure}
 
 \begin{figure}
     \centering
     \includegraphics[scale=0.99]{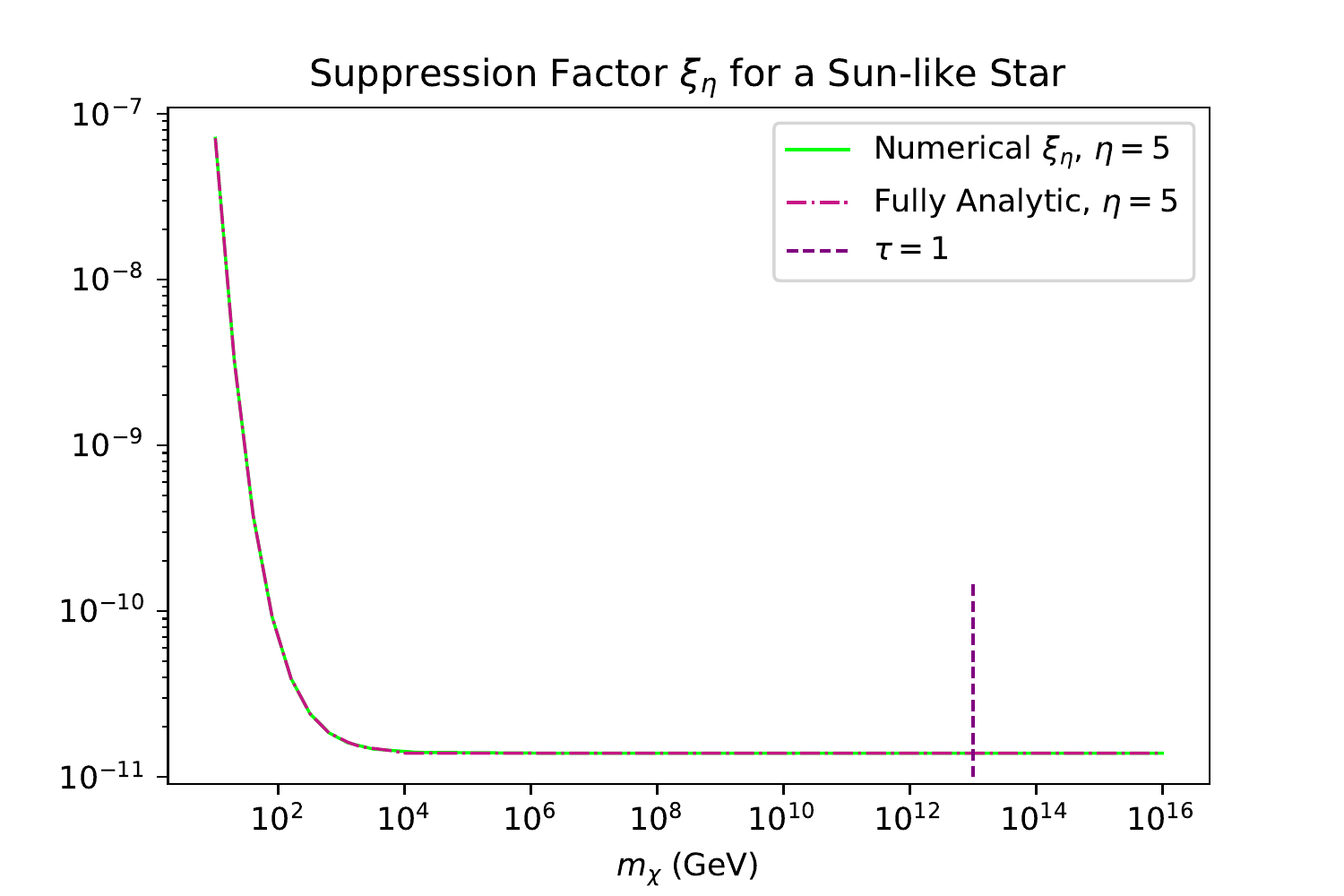}
     \caption{Same as Figure~\ref{fig:SF_sun_full_analytic}, but for sun-like star moving with $\eta=5$. The suppression factor $\xi_{\eta}$ for a sun-like star that has a stellar velocity of approximately $9\times10^7$ cm/s (such that $\eta=5$), much faster than our Sun's velocity relative to its DM halo. This assumes that $\bar{v}=2.2\times10^7$ cm/s, and a DM density of $\rho_{\chi}\approx1$. Note that the suppression is significant in this case, being at least $10^{-7}$, and saturating at $\sim 10^{-11}$.}
     \label{fig:SF_sunlike_full_analytic}
 \end{figure}

 In summary, in this section we have derived and validated simple analytical formulae for the suppression factors of the capture rates in terms of the dimensionless stellar velocity $\eta \equiv \sqrt{\frac{3}{2}}\frac{\tilde{v}}{\bar{v}}$ for both single and multi scatter capture of DM. In the next section we explore the effects of the suppression of the capture rates by stellar velocities in the context of Pop~III stars as DM probes.
 
\section{Bounds on the DM-Nucleon Cross Section from Pop~III stars}\label{sec:Bounds}
 
 In the previous section, we found an analytic closed form for the suppression in capture rate due to the relative velocity between a star and the DM halo. In this section we address the following question: if a Pop~III star does not form precisely at the center of the DM halo, and therefore, has some orbital velocity, how will this affect the constraining power of Pop~III stars on DM parameters such as the DM-nucleon scattering cross section.
 
 In most cases, a Maxwell-Boltzmann velocity distribution has been applied to calculate DM capture rates of Pop~III stars. This is because it is typically assumed that they form at the center of DM halos. As a result, they do not have a velocity relative to the halo. Simulations demonstrate that Pop~III stars would form near the center of DM halos in low multiplicity~\citep{Barkana:2000,Abel:2001,Bromm:2003,Yoshida:2006,Yoshida:2008,Loeb:2010,Bromm:2013,Machida:2013,Klessen:2018}. These stars would orbit around the center of the DM halo. They therefore have a relative velocity directly related to the star's distance from the halo's center. We point out that the formalism developed here, and the analytical approximations, are valid for any DM capturing object, such as stars, neutron stars, and brown dwarfs, and we use Pop~III stars just as an example of how to apply it.    

In order to isolate the effects of the stellar velocity on the capture rate, we first consider the extreme case, where the star forms at the scale radius of the halo. This scenario is highly unlikely, as Pop~III stars form much closer to the center of the DM halo; as such the suppression due to Pop~III stars' stellar velocities is expected to be always less than whatever suppression we will find for this benchmark, overly conservative case.  At first pass we assume that the halo follows a Navarro-Frenk-White (NFW) profile~\citep{Navarro:1997}:

\begin{equation}\label{eq:NFW}
    \rho_{halo}=\frac{\rho_{0}}{\frac{r}{r_s}(1+\frac{r}{r_s})^2},
\end{equation}
where $r$ is the distance from the center and $r_s$ is the scale radius, and for DM mini-halos in which Pop~III stars form, it has a value that ranges between 3 and 300 parsecs. $\rho_0$ is the central density, defined as:

\begin{equation}\label{eq:rho0}
    \rho_0= \frac{200}{3}\frac{c_{vir}^3}{\ln{(1+c_{vir})}-\frac{c_{vir}}{c_{vir}+1}}\rho_c,
\end{equation}
where $c_{vir}$ represents a concentration parameter $c_{vir}=\frac{r_{vir}}{r_s}$ and ranges in value from 1 to 10~\citep{Freese:2008dmdens}. $\rho_c$ is the critical density and depends on the redshift $z$ in accordance with the Friedmann equation.
\begin{figure}[!thb]
    \centering
    \includegraphics[scale=0.99]{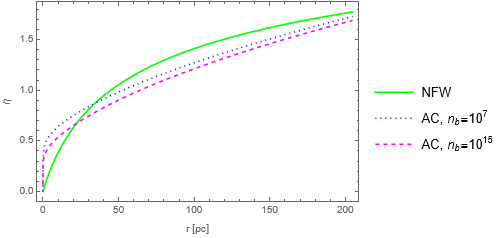}
    \caption{The value of $\eta$ as a function of the distance from the center of a DM halo under various circumstances; the solid green line represents a standard NFW profile, and both the blue (dotted) and pink (dashed) lines are adiabatically contracted (AC). Note that we assume circular orbits. The blue line takes a core density ($n_B$) of $10^7 \text{cm}^{-3}$ and the pink line takes $n_B=10^{15} \text{cm}^{-3}$. The total mass of the halo is $10^6 M_{\odot}$, the redshift is $z=15$, and the concentration parameter is $c=10$. Note that as a result of these parameters, the scale radius is 20.47 pc away from the halo's center. At this location for an NFW profile, $\eta\approx0.69$; for an AC profile with $n_B=10^7$, we get $\eta\approx0.75$; for an AC profile with $n_B=10^{15}$, we get $\eta\approx0.66$. An important takeaway is that, within the scale radius--- the region with which we are concerned--- adiabatic contraction effectively enhances the value of $\eta$ in comparison to the value expected from a standard NFW profile.}
    \label{fig:eta_vs_r}
\end{figure}

Knowing the density distribution of DM in the halo, we calculate the mass enclosed in the stellar orbit, and thus can easily find the speed at which a Pop~III star located at this point would orbit around its center. The stellar velocity will be encoded in a parameter called $\eta$, a dimensionless quantity which is defined as in Equation~(\ref{eq:eta}), which we reproduce here for convenience:  
$$
\eta \equiv \sqrt{\frac{3}{2}}\frac{\tilde{v}}{\bar{v}},
$$
with $\tilde{v}$ representing the stellar velocity and $\bar{v}$ the dispersion velocity of DM. Adopting the parameters described in the caption to Fig \ref{fig:eta_vs_r}, we expect an object located at the scale radius of the halo to have an orbital velocity of $\tilde{v}\approx 5.22 \times 10^5$ cm/s when placed in a standard NFW profile. Of course, changing the redshift or concentration parameter, for instance, would yield slightly different values; we provide an analysis adopting $z=15$ and $c=10$ in order to illustrate one example in depth. Including the effects of the adiabatic compression~\citep{Young:1980,Blumenthal:1985,Freese:2008dmdens,Gendin:2011} on the DM density profile would not affect much this value, since the mass enclosed within the scale radius will stay roughly constant, as the adiabatic compression operates at smaller, sub-parsec scales. 
The relation between the value of $\eta$ and the distance from the halo center is shown in Figure~\ref{fig:eta_vs_r} for a standard NFW profile as well as two adiabatically contracted (AC) profiles. As the baryonic molecular cloud collapses to form a proto-star, the DM orbits respond to this enhancement of the gravitational potential by becoming more tightly packed, a consequence of conservation of adiabatic invariants, such as angular momentum {or radial action}. This is, in essence, what in the literature is called ``adiabatic contraction.'' As commonly done in the literature (see~\cite{Freese:2008dmdens} for example) we use the standard \cite{Blumenthal:1985} formalism to estimate the DM densities. {This formalism assumes circular DM orbits, and, as such, the only relevant adiabatic invariant being angular momentum.}

We elaborate below some of the details of the calculation of the dimensionless stellar velocity $\eta$. The mass profile of the halo is found by integrating over the density profile considered:

\begin{equation}\label{eq:Mhalo_general}
    M(r)=\int_0^{r_{vir}}\rho_{halo}(r)\times 4\pi r^2 dr,
\end{equation}
where, for an NFW profile, we obtain:

\begin{equation}\label{eq:Mhalo_NFW}
    M(r)=4\pi r_s^3 \rho_0 \Biggl[-\frac{r}{(1+ \frac{r}{r_s})r_s}+\ln\Biggl(1+\frac{r}{r_s}\Biggr)\Biggr].
\end{equation}
After substituting the mass profile into $\eta$, we obtain the following expression for a standard NFW profile:

\begin{equation}\label{eq:eta_vs_r_NFW}
   \eta(r) = \frac{\sqrt{\pi }}{50000} \sqrt{\frac{c_{vir}^3 G \text{$\rho_c $} r_s^3 \left(\ln
   \left(\frac{r}{r_s}+1\right)-\frac{r}{r_s \left(\frac{r}{r_s}+1\right)}\right)}{r \left(\ln
   (c_{vir}+1)-\frac{c_{vir}}{c_{vir}+1}\right)}}.
\end{equation}

Knowing the velocity, and the value of $\eta$, for a Pop~III star at a given distance, we can now apply Equation~(\ref{eq:boostMB}). We choose to select the scale radius of the DM halo as a reasonable maximum bound to use when considering boosted capture of Pop~III stars, because in practice, these stars are expected to form well inside the scale radius of DM halos. In turn, this will lead to the highest possible suppression on the previously calculated capture rates in~\cite{Ilie:2019,Ilie:2020BNFa,Ilie:2020PopIIIa}. Note that in our capture rate calculation, since we are mainly focusing on the effect of stellar velocity, we take the assumption that the DM density is fixed, i.e., is the same value at the scale radius as at the halo center.

In order to numerically calculate the capture rate of DM, we need to adopt parameters of Pop~III stars from numerical simulations. Although Pop~III stars are still theoretical objects and have not been observed, simulations have been done, such as for example in~\cite{Iocco:2008, Ohkubo:2009}. In~\cite{Ilie:2019}, it has been shown that Pop~III stars have two different homology scaling relations (in two different mass regimes), where stars with a mass $M_{\star}<20M_{\odot}$ follow $R_{\star}\propto M_{\star}^{0.21}$, and larger mass stars follow $R_{\star}\propto M_{\star}^{0.56}$.

Since our aim in this paper is to understand and quantify the effects of the stellar velocity on DM capture, we assume, for now, the same ambient density at the location of the star, in order to disambiguate between the suppression due to an increase in the stellar velocity, and the decrease in the DM density. Both of those lead to a suppression in the capture rate. For the latter, the effect is trivial, since the total capture rate scales linearly with the DM density: $C_{tot}\sim\rho_{\chi}$. Our aim is to obtain a simple, analytic procedure, that would estimate the suppression rate on capture rates by any astrophysical object, if the parameter $\eta$ is known. 
Previous work in the literature that use compact astrophysical objects as DM probes, typically neglect the effects of the stellar velocity. For example Neutron Stars are considered by~\cite{Bramante:2017}, and exoplanets by~\cite{Leane:2020wob}, and both works neglect the possible role of the relative velocity between the capturing object and the DM halo. The formalism we will develop in Section~\ref{sec:AnalyticalCap} can be easily applied to any such scenario, if the location (and therefore velocity) of the object in question is known. 

\begin{figure}[!thb]
    \centering
    \includegraphics[scale=0.99]{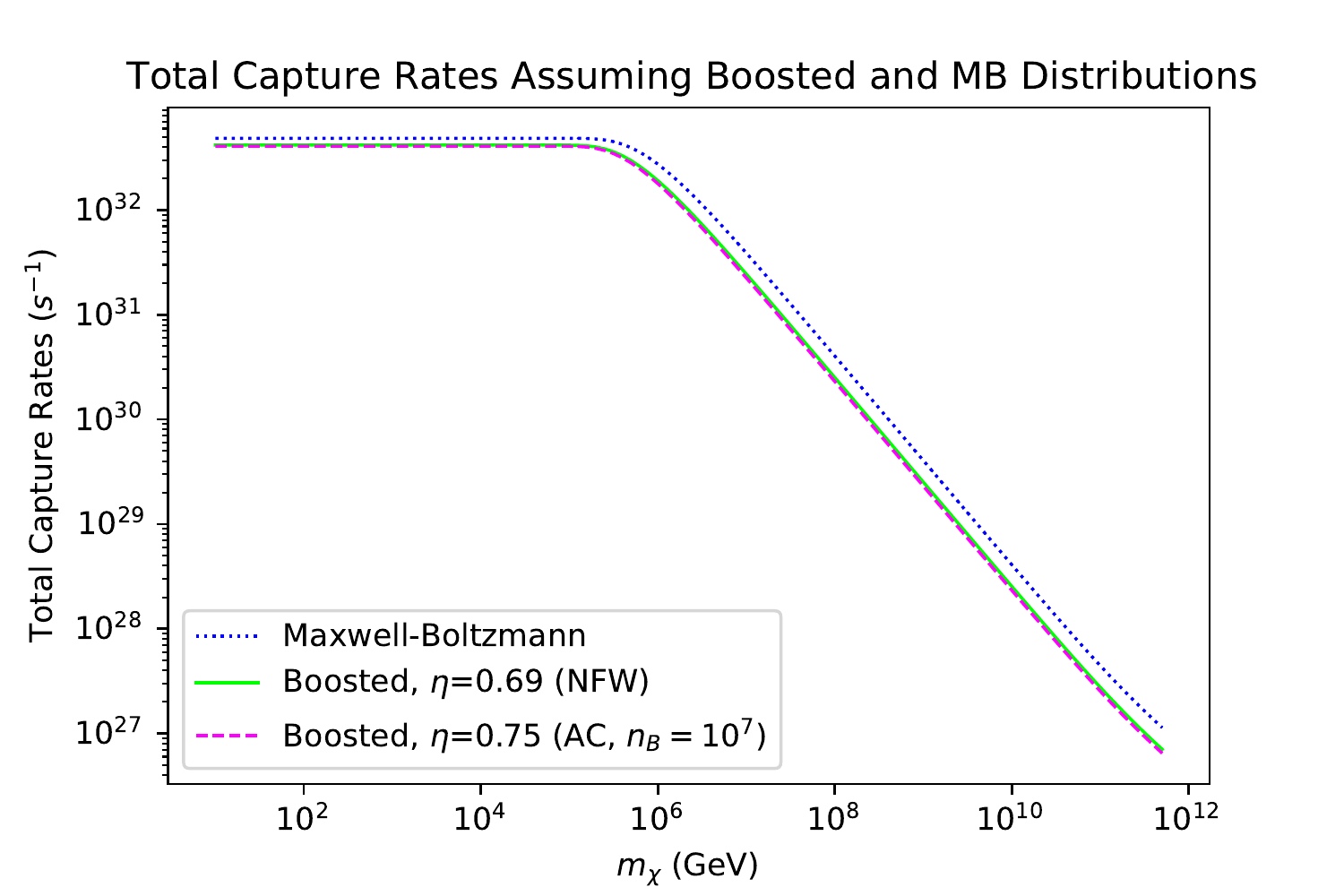}
    \caption{DM capture rates for a 1000 $\Msun$ Pop III star, assuming $\rho_\chi=10^9~\GeV\percc$. The (dotted) blue line shows the total capture rate when assuming a Maxwell-Boltzmann velocity distribution, and the (solid) green and (dashed) pink lines assume a boosted distribution. The green line assumes an NFW profile whereas the pink line is adiabatically contracted with a central density of $10^7~\percc$. Although the capture rates look very close on this graph, the boosted capture rate is actually suppressed by a significant factor. Refer to Figure~\ref{fig:SFPopIII} for a more thorough understanding of the value of this factor.}
    \label{fig:TotalCaptureRates}
\end{figure}

In Fig~\ref{fig:TotalCaptureRates} we contrast the total capture rates of DM by an arbitrary Pop~III star, first placed at the center of the DM halo (as previously assumed) and then placed at the scale radius of the DM halo. We note that, to a good approximation, the capture rates remain unaffected by the inclusion of the stellar velocity, for the case of Pop~III stars. When the boosted distribution is applied, the DM capture rate (see Eq~(\ref{eq:CN})) is suppressed, as one may expect. We next take the ratio between the capture rates calculated using a boosted ($\eta\neq0$) and a regular ($\eta=0$) Maxwell-Boltzmann distribution to illustrate the amount by which capture is suppressed. As shown in Figure~\ref{fig:SFPopIII}, the ratio plateaus for low and high DM masses. Notice that the drastic change in this ratio occurs when the DM mass reaches $10^5$~GeV, which corresponds exactly to the $m_{\chi}$ for which the quantity $k=3\frac{m_T}{m_{\chi}}\frac{v_{esc}^2}{\bar{v}^2}$, defined in Equation~(\ref{eq:Defk}), reaches a value of 1.
\begin{figure}[!thb]
    \centering
    \includegraphics[scale=.99]{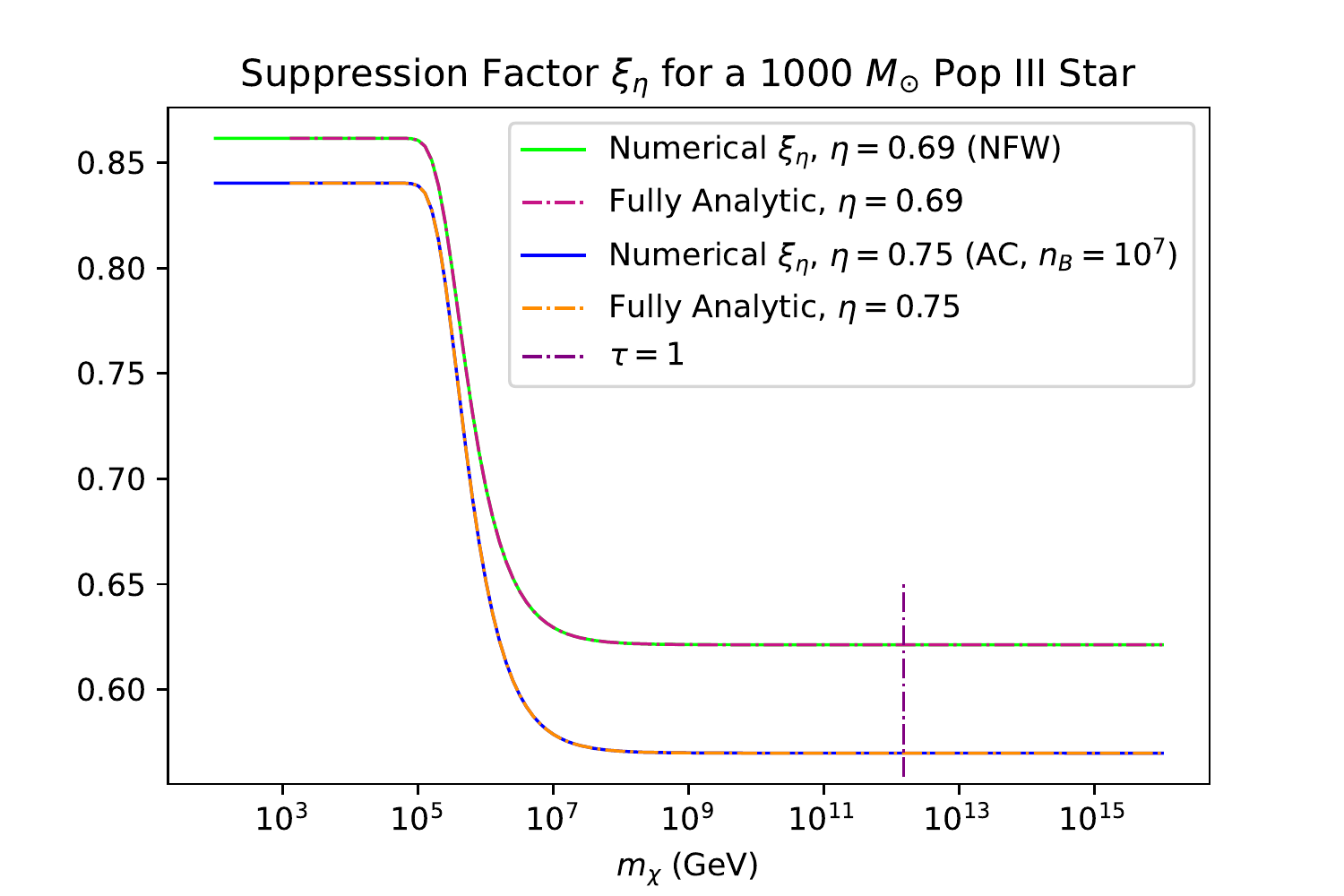}
    \caption{The suppression factor is determined by a numeric calculation (solid lines) or fully analytically (dash dotted lines) via Equations~(\ref{eq:Ieta})-(\ref{eq:I0}). Note the excellent agreement between the two procedures. We consider here a $1000~M_{\odot}$ Pop~III star, with a corresponding radius of $12.85~R_{\odot}$, $\bar{v}=10^6$ cm/s, and $\rho_{\chi}=10^9$ $\text{GeV}/\text{cm}^3$. The dashed purple vertical line corresponds to the transition from single to multiscattering capture, assuming $\sigma$ at the deepest bounds given by X1T~\citep{Aprile:2020}. Note however, that the suppression factor is independent of $\sigma$.}
    \label{fig:SFPopIII}
\end{figure}
 
 Prior works such as~\cite{Ilie:2020BNFa,Ilie:2020PopIIIa} constrain the bounds on the cross section of interaction between DM and baryonic particles due to the impact DM has on the luminosity of Pop~III stars. Any object that is gravitationally bound, such as a star, will have an upper bound on how bright it can shine, at a given mass, i.e., the Eddington limit:
 \begin{equation}\label{eq:subEdd}
     L_{Edd}\leq L_{nuc} + L_{DM},
 \end{equation}
 where $L_{nuc}$ is the luminosity due to nuclear fusion, and $L_{DM}$ is the additional luminosity provided by DM annihilations, which is directly related to the amount of DM captured:
 \begin{equation}\label{eq:LDM}
     L_{DM} =f~C_{tot}~m_{\chi},
 \end{equation}
 where $f$ is the efficiency with which DM annihilation contributes to the luminosity of the star, i.e., the amount of energy thermalized with the star. The remainder $1-f$ is lost to products of annihilation that escape, such as neutrinos. Because $C_{tot}$ is dependent on $\sigma$, we can numerically calculate the maximum expected value of the cross section by finding the maximum value of $L_{DM}$. 
 Recall that the Eddington luminosity is given by
 \begin{equation}\label{eq:LEdd}
     L_{Edd}=\frac{4\pi~ c~G~M_\star}{\kappa_{\rho}},
 \end{equation}
 where $c$ is the speed of light, $G$ is the gravitational constant, $M_{\star}$ is the stellar mass, and $\kappa_{\rho}$ is the opacity of the stellar atmosphere.
 
 The value of $L_{nuc}$ is dependent on the mass of the star ($x\equiv\Mstar/\Msun$), and here we use the fitting form found in~\cite{Ilie:2020PopIIIa}, which we reproduce here for convenience:
 \begin{equation} \label{eq:Lnuc}
L_{n u c} \simeq 10^{\frac{\log \left(3.71 \times 10^{4} L_{\odot} \mathrm{s} / \mathrm{erg}\right)}{1+\exp (-0.85 \log(x)-1.95)}} \cdot x^{\frac{2.01}{x^{0.48}+1}} \operatorname{erg} / \mathrm{s}.
\end{equation}

We note here that the above equation does not take into account the effect DM heating has on the internal structure of the star, specifically on the core temperature that directly affects the nuclear luminosity. Moreover we ignored the DM heating effects on the stellar radius, which in turn affects the capture rate. We have used these models in order to facilitate comparison with earlier work. An accurate calculation requires incorporating DM heating into a stellar structure code, which is beyond the scope of this paper.~\footnote{Such investigations have been performed in the past by~\cite{Iocco:2008}, who find that the hydrogen burning lifetime is prolonged by factors of order of a few, ranging from $5$ for $40~\Msun$ Pop~III stars to $2$ for $600~\Msun$ Pop~III stars. This, in turn, shows the nuclear luminosity for the most massive Pop~III stars is only marginally affected by the effects of captured DM heating.}
\begin{figure}[!ht]
    \centering
    \includegraphics[scale=0.45]{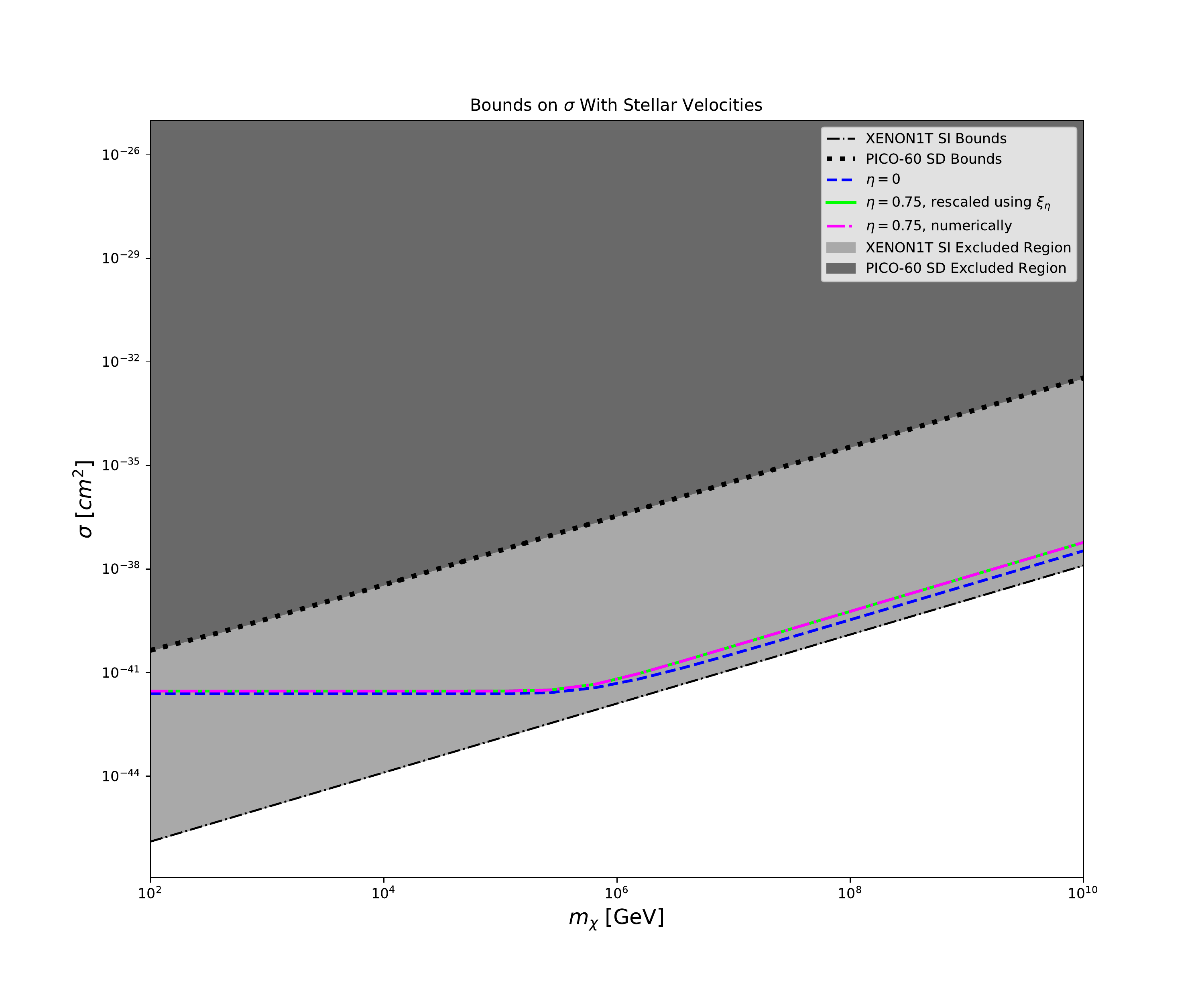}
    \caption{Bounds on the cross section for $\eta=0$ (blue) and $\eta=0.75$ (pink, green) for a 1000 $M_{\odot}$ Pop III star. The DM density taken is $10^{13} ~ \text{GeV cm}^{-3}$. For comparison we add the X1T(SI) and Pico60 (SD) excluded regions.}
    \label{fig:sigmaBounds}
\end{figure}

 Since the capture rate is suppressed, when including the effects of the stellar velocity, the bounds shift upwards by exactly a factor of $\xi_\eta^{-1}$ and become less stringent, as illustrated in Figure \ref{fig:sigmaBounds}. However, note that the values of $\eta=0.69$ and $\eta=0.75$ considered here are, for Pop~III stars, unrealistically high. That is because they correspond to the star at the scale radius of the DM halo, which is many orders of magnitude above the typically expected maximum tens of AU from the center where Pop~III stars form. Even with this exaggerated values of $\eta$ we note that the suppression in the capture rates, and correspondingly the weakening of the cross section bounds are, at most approximately $55\%$. This suppression has a negligible effect on constraints placed on the DM-nucleon cross section, as demonstrated in Fig \ref{fig:sigmaBounds}. There is no significant difference in the bounds on the cross section for all of the values of eta tested. Note that the bounds on $\sigma$ shown here are calculated both numerically assuming a boosted distribution throughout the calculation, and by re-scaling bounds found under the assumption $\eta=0$ by a factor of the inverse of the suppression factor as in $\xi_{\eta}$. We point out that both of these methods produce an exact match (in the figure, the green and pink overlap exactly).  
 
 \section{Conclusion}\label{sec:Conclusions}
 In this paper we derived and validated an analytic closed form of the suppression factor for the capture rates of DM by astrophysical objects that have a non zero velocity with respect to the DM halo: Equations~(\ref{eq:Ieta}) through~(\ref{eq:I0}). One of the most useful applications of those formulae, is that they allow the immediate rescaling of any bounds previously obtained, for any object, under the assumption of zero stellar velocity. Namely, if the stellar velocity is determined, all one needs to do is to rescale the previously obtained bounds $\sigma(\eta=0)$ with the inverse of the suppression factor $\xi_\eta$. For the case of Pop~III stars as DM probes, we find that the role of the stellar velocity can be safely neglected, and all of our previous results, where Pop~III were considered to be at rest with respect to the DM halo, remain largely unchanged. This is because the DM capture rate is suppressed by a factor of $57\%$ at the most. This happens for high mass DM particles ($m_\chi\gtrsim 10^7~\GeV$), and when the star is considered to have formed--- or migrated--- all the way to the scale radius of the DM halo, which is a highly unrealistic scenario.  In most cases Pop~III stars will live much closer to the center of the DM halo, within the inner $10$~AU or so, leading to much higher suppression rates. Of course, the instance of DM halo mergers would change these results, as the location of stars could change significantly in the process. Our formalism is even more relevant for astrophysical objects within the Milky Way that act as DM probes, such as neutron stars, brown dwarfs, exoplanets. In this case, the most promising location in terms of the high DM density, the center of the Milky Way, is the site of a supermassive black hole, which would lead to large orbital velocities, when compared to the center of high redshift DM microhalos, and therefore larger suppression factors for the DM capture. Moreover, for very dim probes, such as neutron stars, the most optimal location would be in the solar system vicinity, where the suppression factor would be even more significant, and therefore important to take into account and estimate. 
 
\begin{acknowledgments}
JP thanks the financial support from Colgate University, via the Research Council student wage grant, and the Justus ’43 and Jayne Schlichting Student Research Funds.
\end{acknowledgments}


\bibliography{RefsDM}{}
\bibliographystyle{aasjournal}



\end{document}